\documentclass[12pt]{spieman}  
\usepackage{amsmath,amsfonts,amssymb}
\usepackage{graphicx}
\usepackage{setspace}
\usepackage{tocloft}
\usepackage{subcaption}

\definecolor{dark-red}{RGB}{238,0,3}
\definecolor{dark-blue}{RGB}{9,71,171}
\definecolor{dark-green}{RGB}{17,149,17}
\definecolor{deep-purple}{RGB}{165,0,220}
\definecolor{dark-yellow}{RGB}{175,175,0}

\title{First results on SiSeRO (Single electron Sensitive Read Out) devices --- a new X-ray detector for scientific instrumentation}

\author[a]{Tanmoy Chattopadhyay}
\author[a]{Sven Herrmann}
\author[b]{Barry Burke}
\author[b]{Kevan Donlon}
\author[c]{Gregory Prigozhin}
\author[a]{R. Glenn Morris}
\author[a]{Peter Orel}
\author[b]{Michael Cooper}
\author[c]{Andrew Malonis}
\author[a]{Dan Wilkins}
\author[b]{Vyshnavi Suntharalingam}
\author[a,d,e]{Steven W. Allen}
\author[c]{Marshall Bautz}
\author[b]{Chris Leitz}

\affil[a]{Kavli Institute of Astrophysics and Cosmology, Stanford University, 452 Lomita Mall, Stanford, CA 94305, USA}
\affil[b]{MIT Lincoln Laboratory, Lexington, MA, USA}
\affil[c]{Kavli Institute for Astrophysics and Space Research, Massachusetts Institute of Technology, Cambridge, MA, USA}
\affil[d]{SLAC National Accelerator Laboratory, 2575 Sand Hill Road, Menlo Park, CA 94025, USA}
\affil[e]{Department of Physics, Stanford University, 382 Via Pueblo Mall, Stanford CA 94305, USA}

\cftpagenumbersoff{figure}
\cftpagenumbersoff{table} 
\begin{document} 
\maketitle

\begin{abstract}
We present an evaluation of a novel on-chip charge detector, called the Single electron Sensitive Read Out (SiSeRO), for charge-coupled device (CCD) image sensor applications. It uses a p-MOSFET transistor at the output stage with a depleted internal gate beneath the p-MOSFET. Charge transferred to the internal gate modulates the source-drain current of the transistor. We have developed a drain current readout module to characterize the detector. The prototype sensor achieves a charge/current conversion gain of 700 pA per electron, an equivalent noise charge (ENC) of 15 electrons (e$^-$) root mean square (RMS), and a full width half maximum (FWHM) of 230 eV at 5.9 keV. In this paper, we discuss the SiSeRO working principle, the readout module developed at Stanford, and the first characterization test results of the SiSeRO prototypes. While at present only a proof-of-concept experiment, in the near future we plan to use next generation sensors with improved noise performance and an enhanced readout module. In particular, we are developing a readout module enabling Repetitive Non-Destructive Readout (RNDR) of the charge, which can in principle yield sub-electron ENC performance. With these developments, we eventually plan to build a matrix of SiSeRO amplifiers to develop an active pixel sensor with an on-chip ASIC-based readout system. Such a system, with fast readout speeds and sub-electron noise, could be effectively utilized in scientific applications requiring fast and low-noise spectro-imagers.    
\end{abstract}


{\noindent \footnotesize\textbf{*}Tanmoy Chattopadhyay,  \linkable{tanmoyc@stanford.edu} }

\begin{spacing}{1}   

\section{Introduction}
\label{sect:intro}  \label{sec_intro}

Charge-coupled devices (CCDs) have been the detector technology of choice in X-ray astronomy for almost 30 years. First flown on the Advanced Satellite for Cosmology and Astrophysics (ASCA \cite{Tanaka94}) in 1993, these detectors have since been used on many missions, including the flagship Chandra X-ray Observatory\footnote{https://chandra.harvard.edu/} and XMM-Newton\footnote{https://www.cosmos.esa.int/web/xmm-newton} satellites. The next generation of improved CCD detectors are proposed to form a key component of the anticipated instrument suite of future mission concepts including Lynx\footnote{https://www.lynxobservatory.com/}, AXIS\footnote{https://axis.astro.umd.edu/}, and ARCUS\footnote{http://www.arcusxray.org/}. Here the primary benefit of CCDs is their ability to provide good time resolution, low-noise, high spatial, and moderate spectral resolution imaging over relatively large focal planes \cite{Lesser15_ccd,gruner02_ccd}. The CCD detectors aboard Chandra and XMM-Newton have also been shown to operate robustly in space environments for more than 20 years. 

While Silicon-based X-ray imagers routinely show excellent performance, and regularly demonstrate an ENC (equivalent noise charge) of $<$4e$^-$, the application case of small pixels (enhanced charge sharing; supporting the high spatial resolution of concepts like Lynx and AXIS) and soft X-rays ($<$500 eV; an energy range of particular interest for all of these mission concepts)  would greatly benefit from improved noise performance, preferably to the point where single electrons can be counted with confidence. 
MIT Lincoln Laboratory has matured their X-ray CCD technology substantially in recent years \cite{bautz18,bautz19,bautz20}, including the development of a novel Single electron Sensitive Read Output stage (hereinafter SiSeRO) for proof-of-principle measurements. This amplifier draws on earlier work on floating-gate amplifiers that demonstrated extremely high responsivity and sub-electron \cite{matsunaga91} noise, and is similar in some respects to the successful DEPFET sensor \cite{kemmer87_depfet,strueder00_depfet_imager} developed for the Athena wide-field imager \cite{treberspurg20_wfi}. The SiSeRO concept could in principle improve the noise performance of X-ray CCDs by a factor of 10, and provide sub-electron noise for megapixel per second readout rates. 

In this article, we report the first test results of prototype SiSeROs using readout electronics developed at Stanford University. 
The working principle of the SiSeROs is demonstrated in Sec.~\ref{sec_sisero}, followed by a description of the readout module and the test stand in Sec.~\ref{sec_stand}. Characterization test results of a prototype SiSeRO are discussed in Sec.~\ref{sect:sections}. We summarize the results and the future plans for testing the SiSeROs in Sec.~\ref{sec_future}.  

\section{The SiSeRO Principle and Operation}\label{sec_sisero}
\begin{figure}
    \centering
    \begin{subfigure}{.59\textwidth}
  \includegraphics[width=\linewidth]{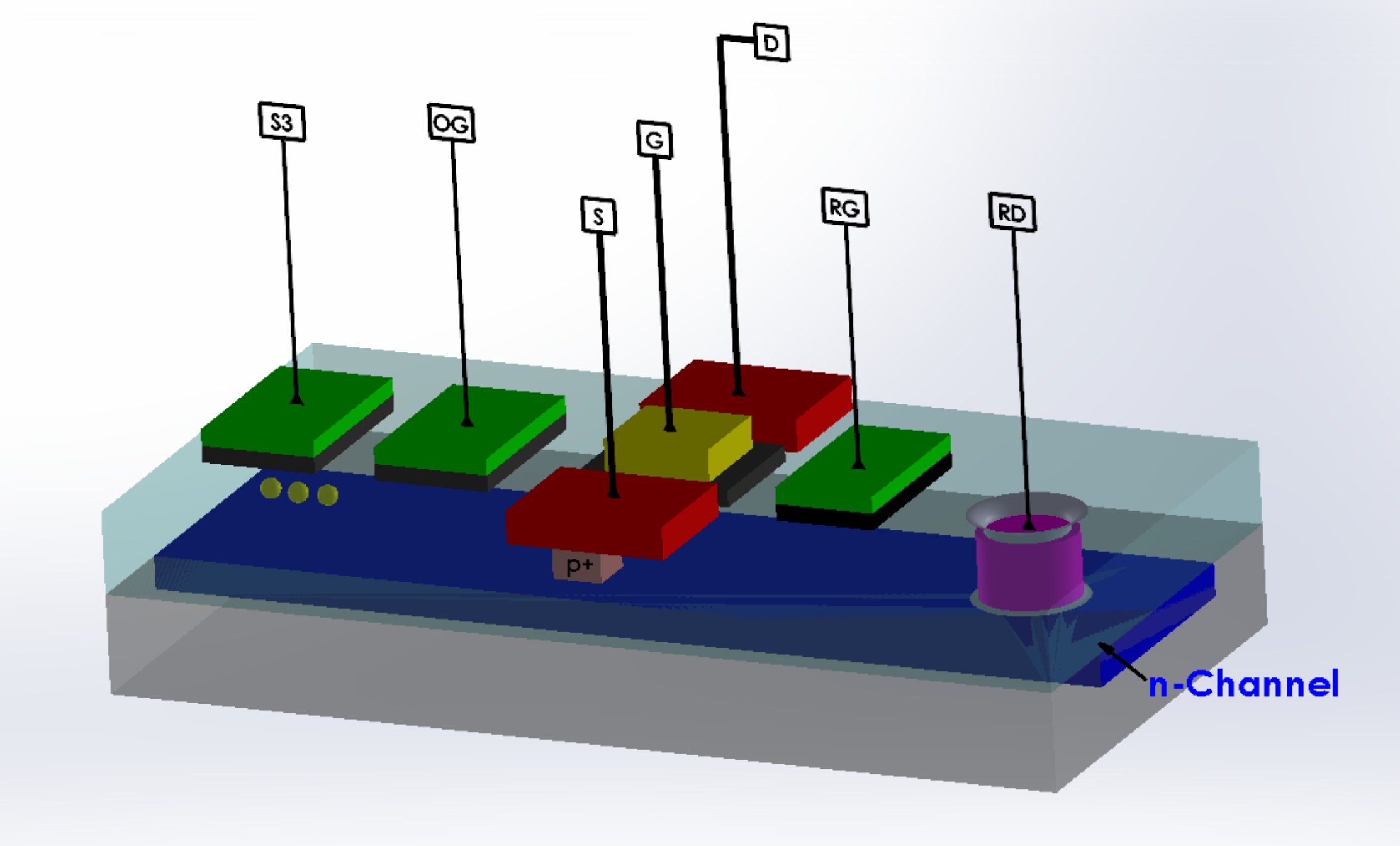}
    \caption{}
    \end{subfigure}
  \begin{subfigure}{.36\textwidth}
     \includegraphics[width=\linewidth]{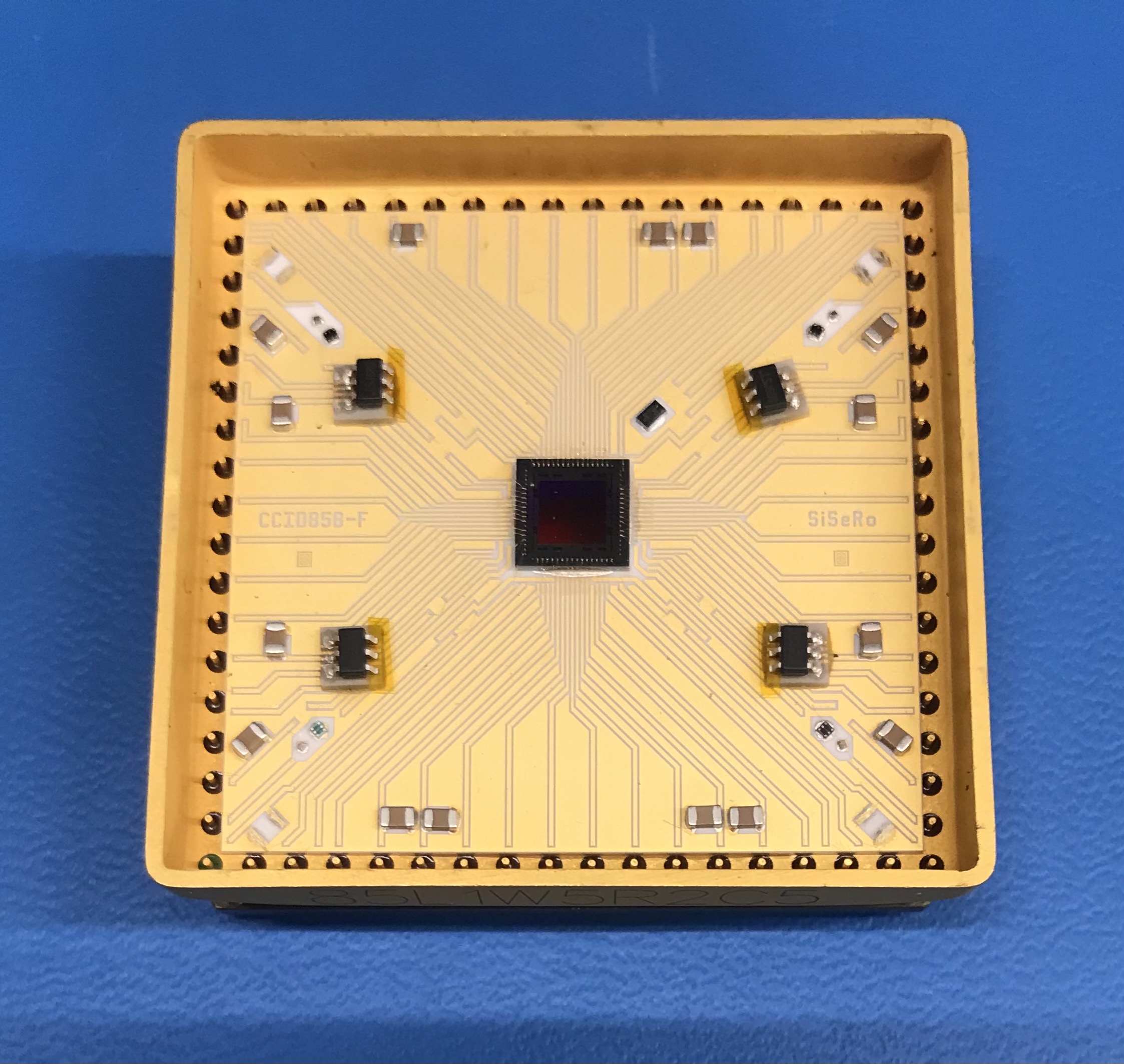}
    \caption{} 
  \end{subfigure}%
    \caption{Working principle of the SiSeRO. (a) Schematic of the SiSeRO. It uses a p-MOSFET transistor with an internal gate beneath the p-MOSFET. Charge transferred from the output gate (OG) to the internal gate modulates the drain current of the p-MOSFET. See text for more details. (b) One SiSeRO in its package (CCID85F), as used for testing.}
    \label{sisero_fig}
\end{figure}

A simple schematic of our prototype SiSeRO device is shown in Fig.~\ref{sisero_fig}a.
It consists of a p-MOSFET transistor that straddles the CCD channel. As the CCD charge packet is transferred beneath the p-MOSFET, it modulates the source-drain current ($I_{\mathrm{DS}}$) of the transistor, with the degree of modulation directly proportional to the transferred charge. 
The p-MOSFET external gate is set to a DC bias and is used to adjust the current through the p-MOSFET to an optimum signal-to-noise (SNR) condition. 

Figure \ref{sisero_fig}b shows the detector package that was used for testing. In the prototype SiSeROs manufactured by MIT Lincoln Laboratory (MITL), 
the transistor and the channel beneath are small, around 2 $\mu$m in width.
The test devices are available primarily in two packages, denoted as CCID85E and CCID85F, each having four different variants of SiSeRO positioned at the four corners of the package. The variants can have different gate structures of the p-MOSFET and different trough locations within the internal channel. The trough implants are introduced in order to have better confinement of the charge packet in the internal channel. The troughs are rectangular with width 0.4 $\mu$m, and can be located either at the center of the channel or towards the source or drain of the transistor. 
The detectors are fabricated in an n-channel CCD, using a low-voltage, single-poly process with two metal layers \cite{bautz19}. The imaging area of the detector is $\sim$ 4 mm $\times$ 4 mm in size, with a 512 $\times$ 512 array of pixels (nominal pixel size 8 $\mu$m). The prototype SiSeRO detectors are currently front-illuminated devices. In Fig.~\ref{sisero_fig}a, S$_3$ and OG  stand for the third (or last) serial clock at the output stage, and the output gate, respectively. OG is set to a DC bias to move the charge from S$_3$ to beneath the p-MOSFET. The charge packet is eventually drained to reset drain (RD), which is set to a high positive DC bias, using a reset clock (RG).
On the packaged devices, the source and drain of the amplifier are brought directly out of the package. This enables the device to be operated in a drain- (or source-) current readout mode where the drain (or source) current is read out by an external readout electronics module. For our testing, we characterized two different SiSeROs: one has the trough implant closer to the drain, the other one has the trough implant located at the center of the CCD channel.  

\section{Characterization Test Stand and Readout Module}
\label{sec_stand}
Here we give a brief description of the characterization test stand, followed by a discussion on the readout module developed to characterize the SiSeROs. 
The same test stand has been used earlier in our laboratory to characterize fast, low-noise X-ray CCDs for astronomical applications. The details of the test stand and its components are discussed in \cite{Chattopadhyay20_spie}. Here we give only a brief description of the setup.

\begin{figure}
    \centering
    \includegraphics[width=.65\linewidth]{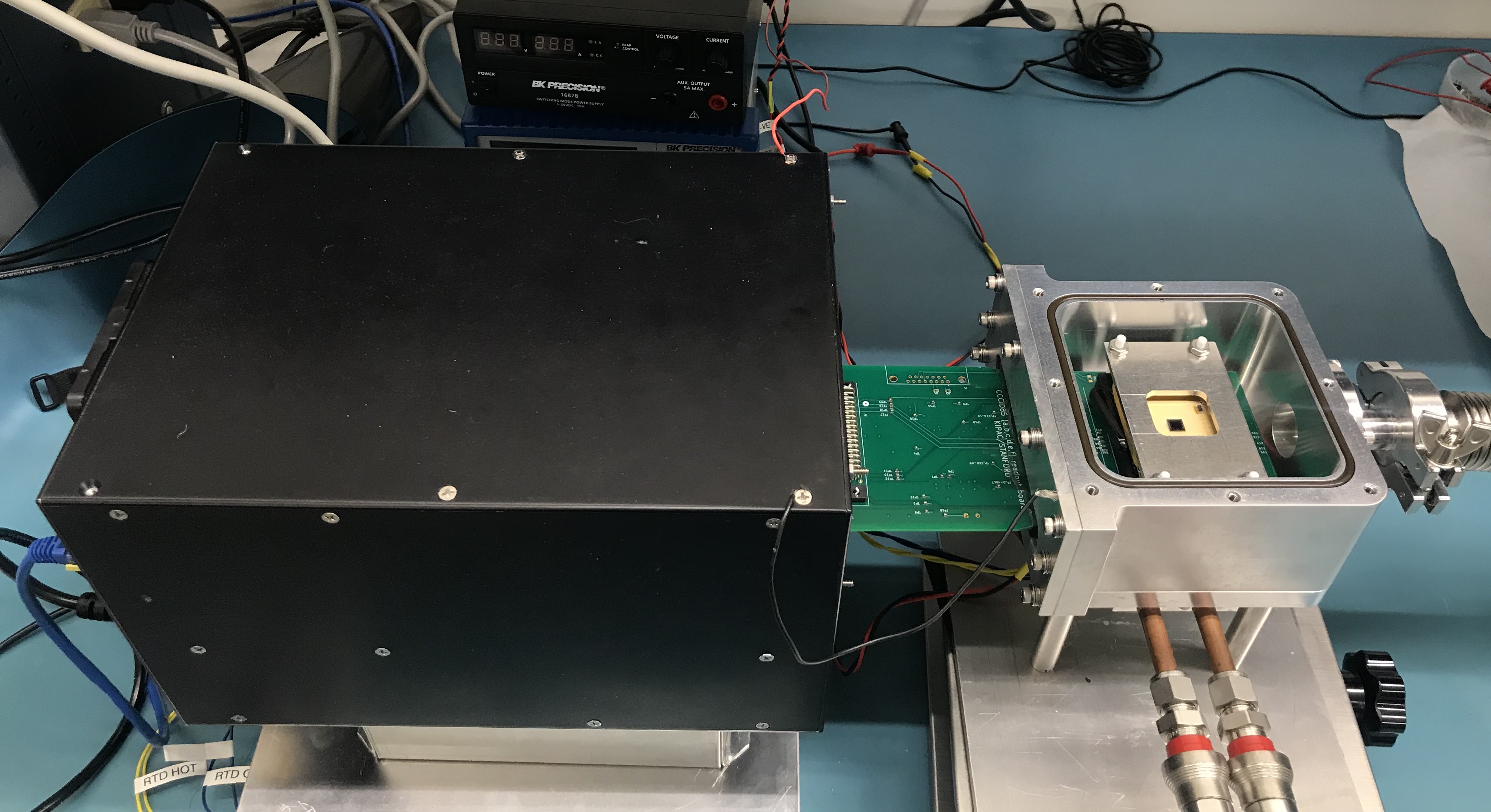}
    \caption{The characterization test stand with a SiSeRO (CCID85F) mounted inside the chamber. The detector is clamped against the thermal cold block with a metal plate from the top for cooling. An interface board routes the clock and bias signals from the Archon controller (black box) to the preamplifier board, and the detector analog output from the board to the Archon differential ADCs. For more details of the test setup and its components, see \cite{Chattopadhyay20_spie}.}
    \label{expt_setup}
    \end{figure}

The experimental set-up is shown in Fig.~\ref{expt_setup}. 
The detector housing is compact in size (13 cm $\times$ 15 cm $\times$ 6.5 cm).
A narrow slot in the side flange is used to epoxy the preamplifier board. There is a mounting socket (a 19 $\times$ 19 position PGA ZIF socket from 3M) for the detector on the vacuum side of the board. 
We cut out a square section from the central region of the socket such that the backside of the device package is in direct thermal contact with an aluminum cold block. The detector is clamped against it with a metal plate from the top to improve the thermal contact.
A thermo-electric cooler (TEC), installed inside the chamber, is used to cool down the device package. A proportional–integral–derivative (PID) controller loop is employed to adjust the current of the TEC and thereby control the temperature of the device with an accuracy better than 0.2$^\circ$C. A water-cooled plate at the bottom flange dissipates the heat from the hot side of the TEC.
There is a small X-ray entrance window, made of a 500 $\mu$m  thick beryllium disc (95 \% transmission of 5.9 keV photons) in the top flange directly above the detector, to allow X-ray photons from the calibration source to characterize the detector. 
The other side of the board is connected to an Archon CCD controller\footnote{http://www.sta-inc.net/archon/} (the black box in Fig.~\ref{expt_setup}). The controller will be described in more detail below.

\subsection{Readout Module}

The readout module is composed of two blocks: a preamplifier that reads the SiSeRO drain current and converts it to a fully differential analog voltage signal; and the Archon controller digitizer, which samples, digitizes the fully differential analog voltage signal, and estimates the charge signal amplitude for each pixel. Below we discuss the preamplifier circuit and the simulation results, followed by a brief description of the Archon controller.   

\begin{figure}
    \centering
    \begin{subfigure}{1\textwidth}
    \includegraphics[width=1.0\linewidth]{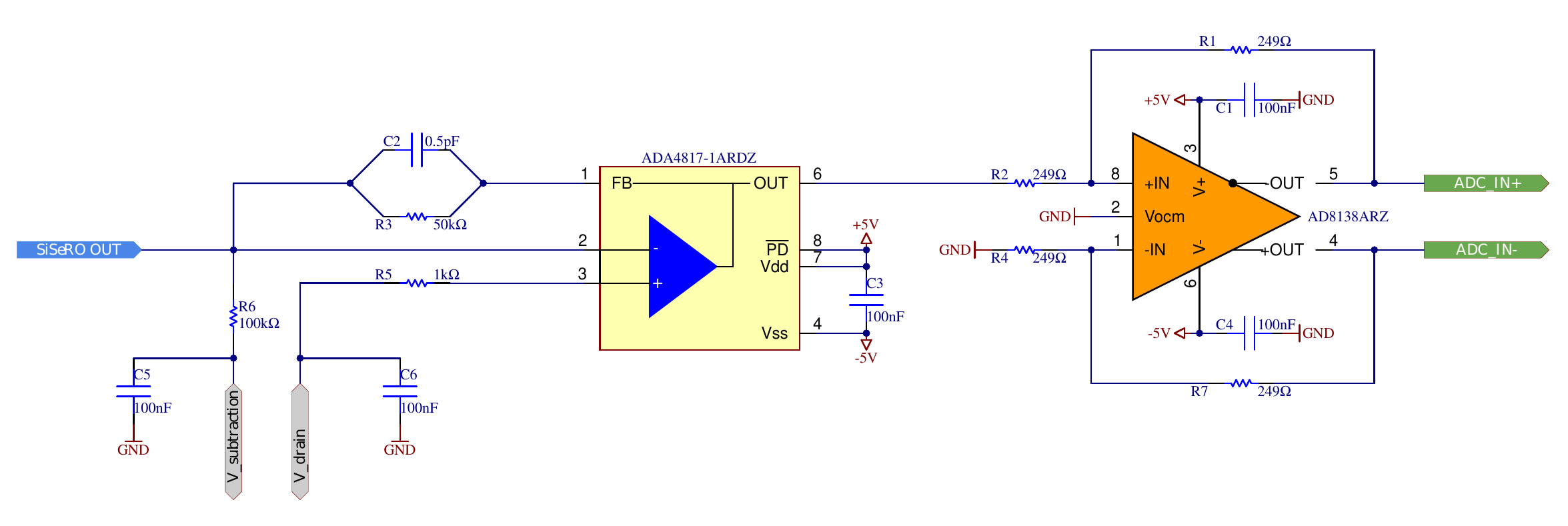}
    \caption{}
    \end{subfigure}
  \begin{subfigure}{.5\textwidth}
     \includegraphics[width=\linewidth]{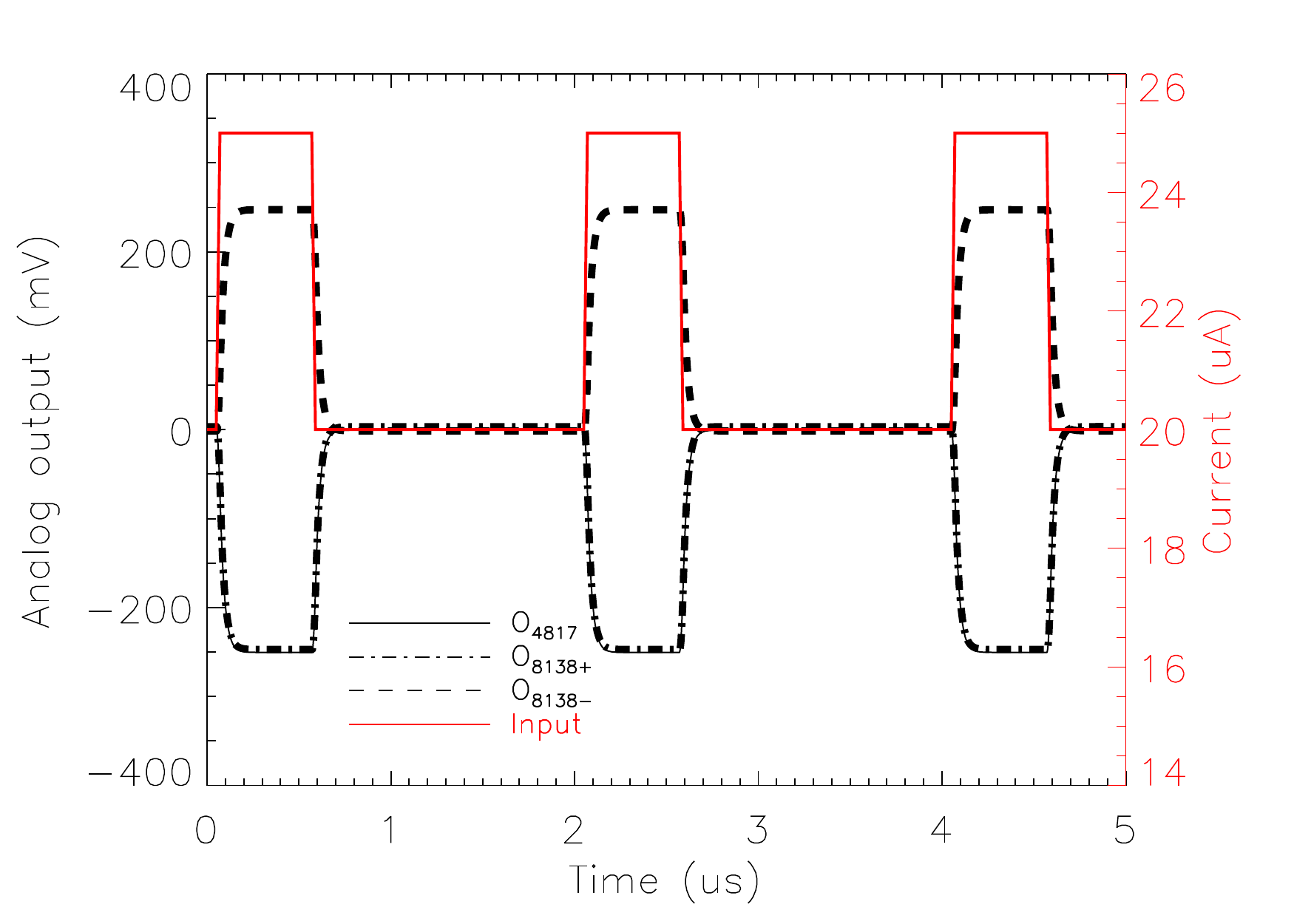}
    \caption{} 
  \end{subfigure}%
  \begin{subfigure}{.5\textwidth}
     \includegraphics[width=\linewidth]{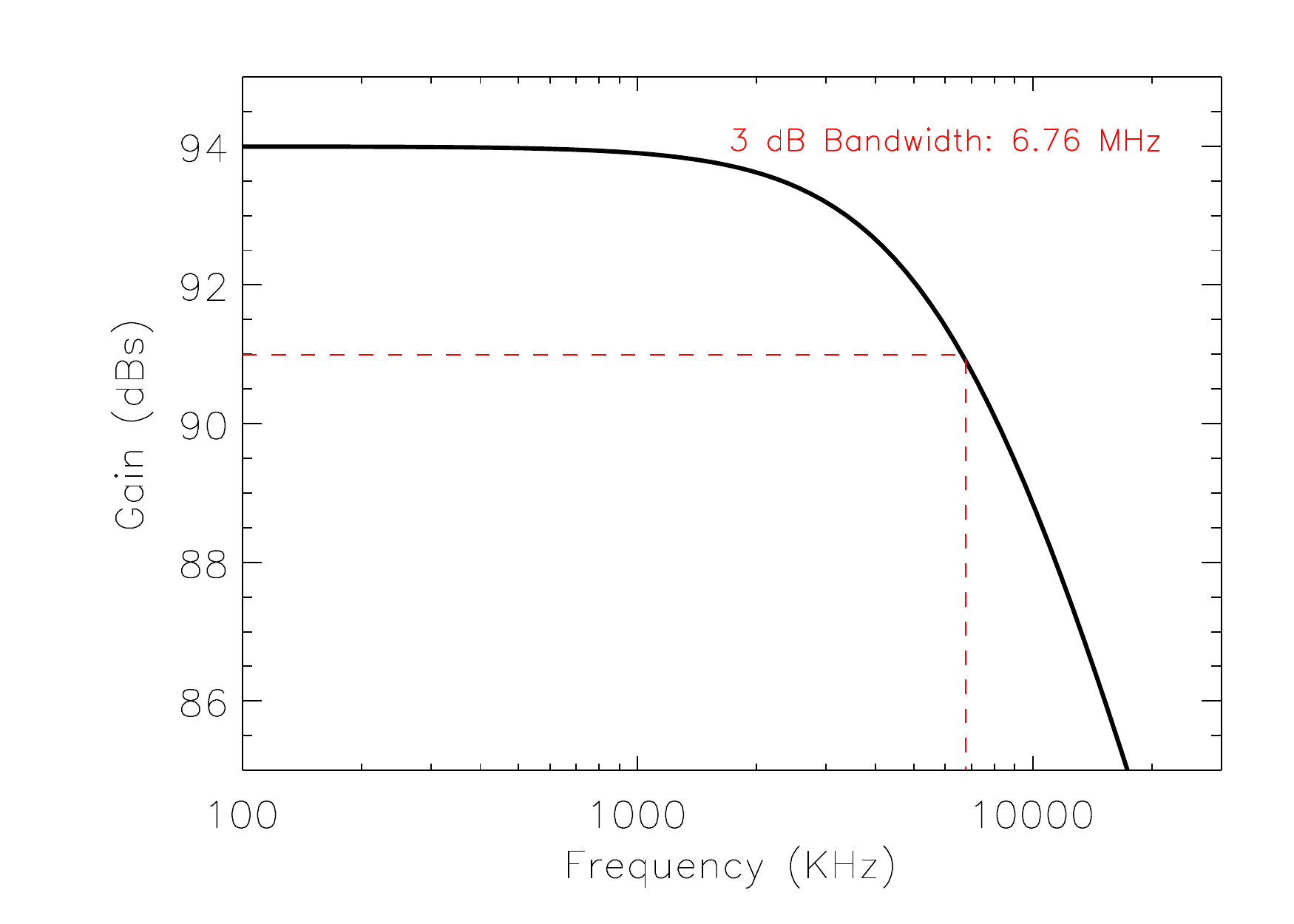}
    \caption{} 
  \end{subfigure}%
    \caption{(a) Schematic of the preamplifier circuit to read out the drain current from the SiSeRO output. It uses an I2V amplifier at the first stage with a single ended operational amplifier (ADA4817). We bias the p-MOSFET drain ($V_{\mathrm{drain}}$) from the non-inverting input of the amplifier. $V_{\mathrm{subtraction}}$ is used to minimize the offset in the analog output of the I2V due to the p-MOSFET current and the input bias current of ADA4817. This stage is followed by a fully differential ADC driver (AD8138). (b) Transient simulation results for the preamplifier chain in LTspice simulator. Outputs from different stages of the preamplifier circuit for a 500 kHz pulsed input current signal (red) are shown as black lines of different styles: solid --- output of ADA4817; dashed --- inverting output of AD8138; dashed-dot --- non-inverting output of AD8138. (c) Predicted bandwidth for the preamplifier circuit. The board is expected to provide $\sim$6.8 MHz of bandwidth.}
    \label{preamp_schematic}
\end{figure}

Figure \ref{preamp_schematic}a shows the schematic of the preamplifier circuit. 
It has two stages: an I2V amplifier to convert the SiSeRO drain current to an analog voltage signal, followed by a differential ADC driver to produce a fully differential output voltage signal. 
For the I2V stage, we use a single-ended operational amplifier (an ADA4817 from Analog Devices\footnote{https://www.analog.com/en/products/ada4817-1.html$\#$product-overview}).
The drain of the p-MOSFET is connected to the inverting input of the amplifier. The voltage at the output of the amplifier is proportional to the SiSeRO drain current $\sim I_{\mathrm{drain}}~R3$, where `$R3$' is the feedback resistor. We provide a DC bias to the drain of the p-MOSFET using $V_{\mathrm{drain}}$ through the non-inverting input of the ADA4817. $V_{\mathrm{subtraction}}$ is a DC bias connected to the SiSeRO input that is used to add or subtract excess current, to adjust the output voltage offset within the allowed voltage swing of the ADA4817 (-2.5--2.5 V).   
In the second stage, for the differential ADC driver, we use an AD8138 from Analog Devices\footnote{https://www.analog.com/en/products/ad8138.html$\#$product-overview}. A gain of unity is implemented in this stage. At the output, we get a differential analog signal, IN+ and IN-, which is fed to the ADC driver of the Archon controller. 
Although our primary focus here is to demonstrate that SiSeROs can be read out in drain current readout mode with reasonable noise and spectral performance, the amplifiers in the circuit design were chosen such that both the amplifiers have large bandwidths, in order to support high readout speeds. The ADA4817 has extremely small input capacitance, resulting in a low RC time constant for the circuit. However, due to the large parasitics of the PCB setup, the overall bandwidth in the current design is limited by the compensation of the I2V to around 6 MHz. The ADA4817 and the AD8138 both feature low voltage noise densities ($<$5 nV / $\sqrt{\mathrm{Hz}}$), which ensures the low noise performance of the setup. 

We simulated the preamplifier circuit using the LTspice simulator\footnote{https://www.analog.com/en/design-center/design-tools-and-calculators/ltspice-simulator.html}. Figure \ref{preamp_schematic}b shows the results of a transient simulation for a 500 kHz pulsed input current signal (shown in red). The output from the I2V amplifier is shown as a solid black line, with the inverting and non-inverting outputs of the ADC driver shown as dashed black and dashed-dot black lines, respectively. 
The circuit bandwidth of 6.7 MHz has been obtained by running a small-signal AC simulation (shown in Fig.~\ref{preamp_schematic}c). The bandwidth is limited by the RC time constant of the I2V amplifier feedback, whose selected values are needed for stability ($1/2\pi R C \approx 6.4$ MHz). 
We also performed voltage noise simulations of the circuit in LTspice. The simulation results indicate an input referred current noise density of around 1.5 pA/$\sqrt{\mathrm{Hz}}$. Consequently, the integrated noise floor is 1.5 nA in a 1 MHz bandwidth, which is also equivalent to an ENC of $\sim 2 \mathrm{e}^{-}$. 

The preamplifier board is connected to the Archon controller through an interface board that routes the clock and biases from the internal Archon modules to the detector, while at the same time the detector analog outputs from the preamplifier board are routed to the differential ADC inputs on the Archon controller. 

The Archon controller, procured from Semiconductor Technology Associates, Inc (STA), is an FPGA-based CCD controller, which provides the necessary bias (in the range -14 -- 31 V) and clocks (14-bit 100 MHz) to the detector, digitizes the detector outputs (using 16-bit 100 MHz ADCs), and estimates the charge signal for each pixel to generate the image.
The Archon receives configuration information about the detector bias, clocking sequence, and sampling of digitized waveforms from a host PC and then returns image data via a gigabit Ethernet connection. 
More details on the Archon controller can be found in \cite{archon14}.
The total (current) signal gain of the implemented drain readout chain is 376 analog digital units (ADU) per $\mu$A of input current.

\section{Characterization of the SiSeROs}
\label{sect:sections}
We characterized two variants of SiSeROs --- one with the trough implant located at the center of the channel, and the other where the trough implant is located closer towards the drain. Understanding the effect of the trough and its location on the charge transfer is essential to mature the design of SiSeROs in the future.  
The device was cooled down to 250 K (-23$^\circ$C) to minimize leakage current. 
We characterized the detector at 500 kilo-pixel per second read out rate. 
The clocking sequence (and thereby the readout speed) of the serial register clocks is defined in a timing script inside the Archon Configuration File (ACF).
The charge packet is first transferred from the imaging region to the serial register of the detector using a three electrode imaging clock sequence (at 400 kHz). The charge in the serial register is moved across the serial gates, S$_1$ to S$_2$ and S$_2$ to S$_3$ sequentially, to the output gate, and finally to the depleted internal channel beneath the p-MOSFET. After the charge is read out, a reset clock (RG) drains the charge packet to a reset drain (RD) kept at high DC potential.  

\begin{figure}
    \centering
    \begin{subfigure}{.49\textwidth}
     \includegraphics[width=1\linewidth]{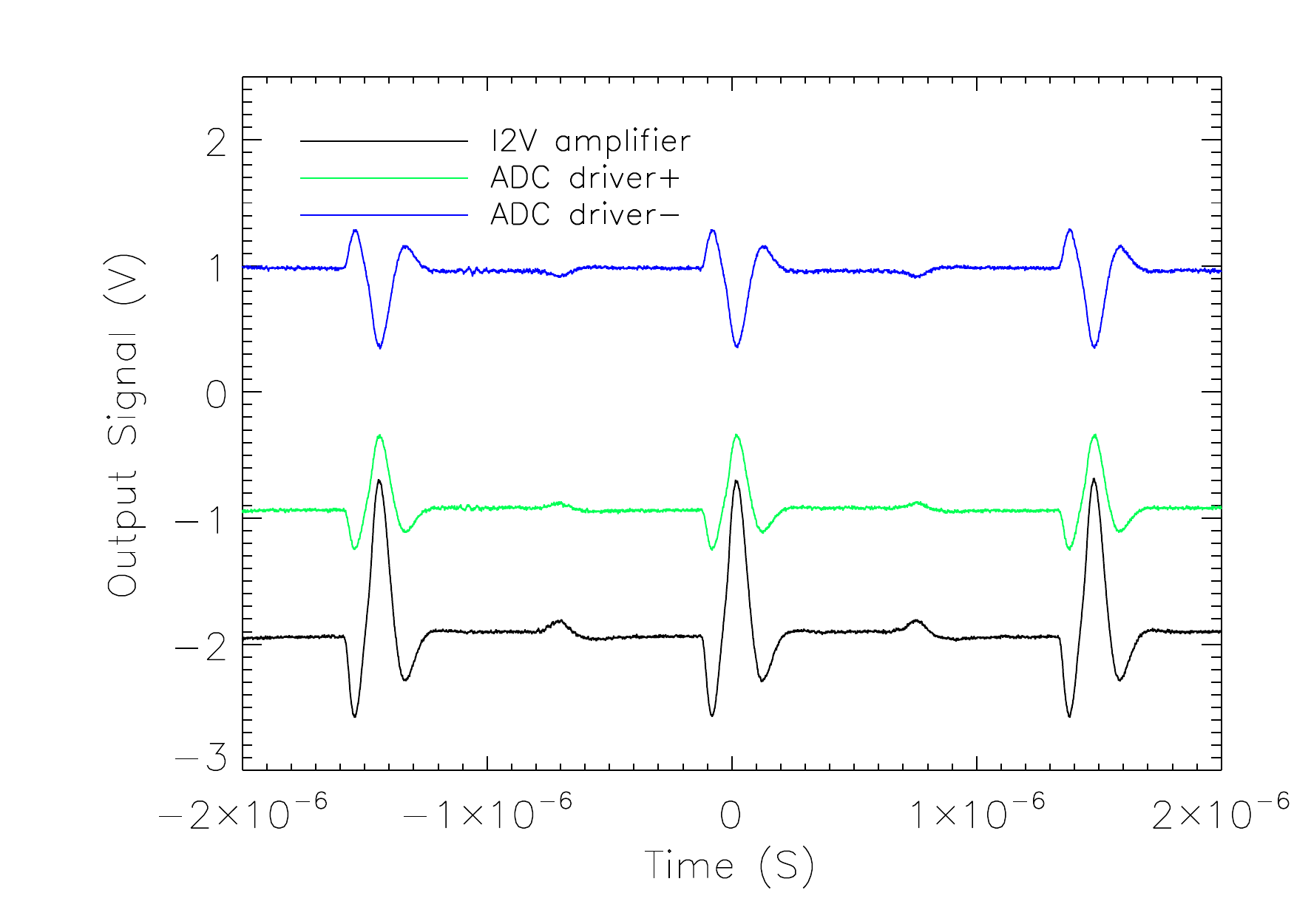}
    \caption{}
    \end{subfigure}
    \begin{subfigure}{.49\textwidth}
    \includegraphics[width=1\linewidth]{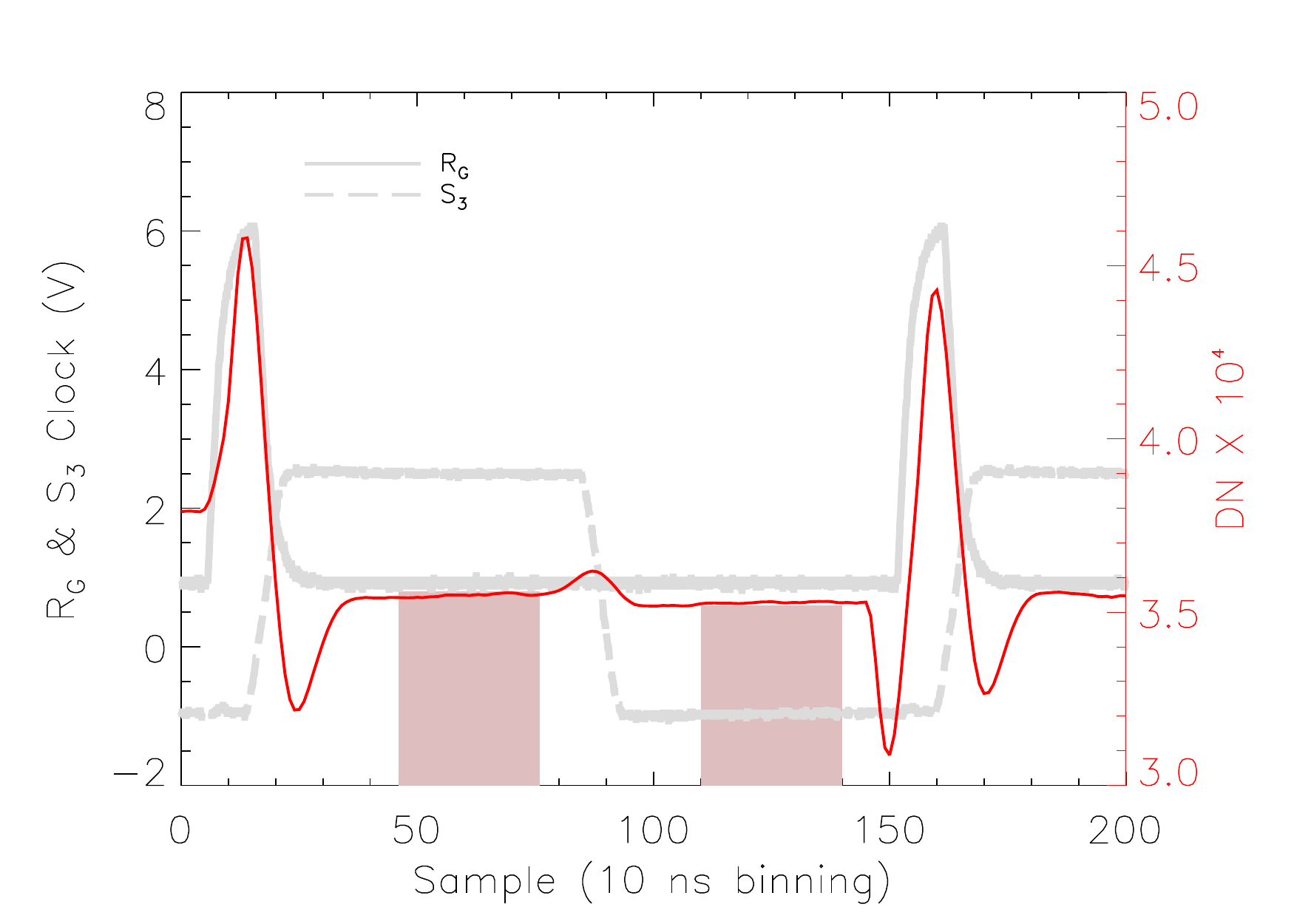}
    \caption{}
    \end{subfigure}
    \caption{(a) Oscilloscope output from the 1$^{st}$ stage I2V amplifier (black) and the ADC driver (green and blue). 
    (b) The output waveform (red solid line plotted against the right axis) obtained from the Archon controller along with the reset, R$_{\mathrm{g}}$ (solid gray line) and S$_3$ (dashed gray line) clocks. The shaded regions represent the baseline (S$_3$ is high, internal gate empty) and video (S$_3$ is low, charge in internal gate) signals which are used for CDS filtering. The video signal starts with charge transfer from S$_3$ (when S$_3$ is low) to the internal gate through the output gate. The reset plateau region starts when R$_{\mathrm{g}}$ sets to baseline.}
    \label{expt_amplifier}
\end{figure}

Figure \ref{expt_amplifier}a shows an oscilloscope output from different stages of the readout module.
The black line is the analog output of the I2V amplifier. The green and the blue lines are the non-inverting and inverting outputs of the differential ADC driver (AD8138), respectively. The readout speed is set at 500 kHz. Each pixel in the detector generates this type of signal at the output. Figure \ref{expt_amplifier}b shows the digitized waveform from the Archon controller as a solid red line. The reset (RG) and S$_3$ clocks, synchronized with the waveform, are shown in solid gray and dashed gray lines, respectively.
The waveform can be seen to have three distinct features. It starts with a reset of the internal channel, shown by a large spike in the signal. Following the reset, the output settles to a baseline level. A charge packet is transferred from S$_3$ to the internal channel through the output gate OG when S$_3$ is low. This can be seen as a small peak in the signal. The difference in the signal levels before and after the charge transfer is proportional to the transferred charge. A correlated double sampling (CDS) function in the Archon software calculates the charge for each pixel from the two signal levels to generate the detector images.      
Image cleaning and processing, event selection, generation of spectra, and estimation of the noise and spectral resolution are carried out using IDL (Interactive Data Language) event processing software.

\subsection{Optimization in Biasing Parameters}

We characterized a SiSeRO p-MOSFET transistor to understand its behavior and optimize biasing conditions for the transistor Source (S), Drain (D), and Gate (G). Fine tuning the bias of the output gate (OG), reset drain (RD), and the reset clock (RG Low and RG High) is important to assure complete charge transfer to the internal channel, and that there is no back injection of charge to the channel. We used an off-centered SiSeRO for detailed characterization. Since the transistors in all the SiSeRO variants are identical, we expect similar I-V behavior for them.           

\begin{figure}
    \centering
    \begin{subfigure}{.49\textwidth}
     \includegraphics[width=1\linewidth]{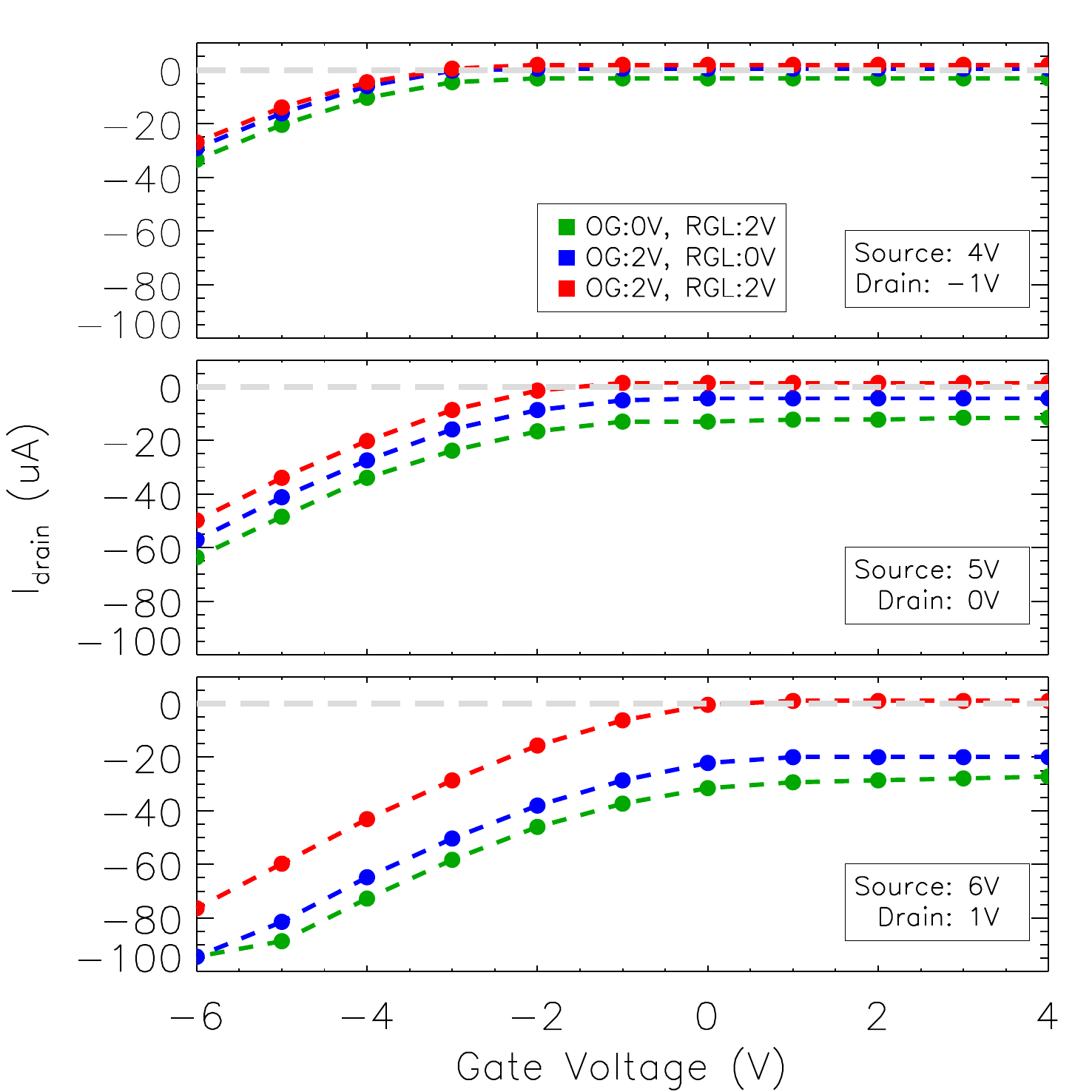}
    \caption{}
    \end{subfigure}
  \begin{subfigure}{.5\textwidth}
     \includegraphics[width=1\linewidth]{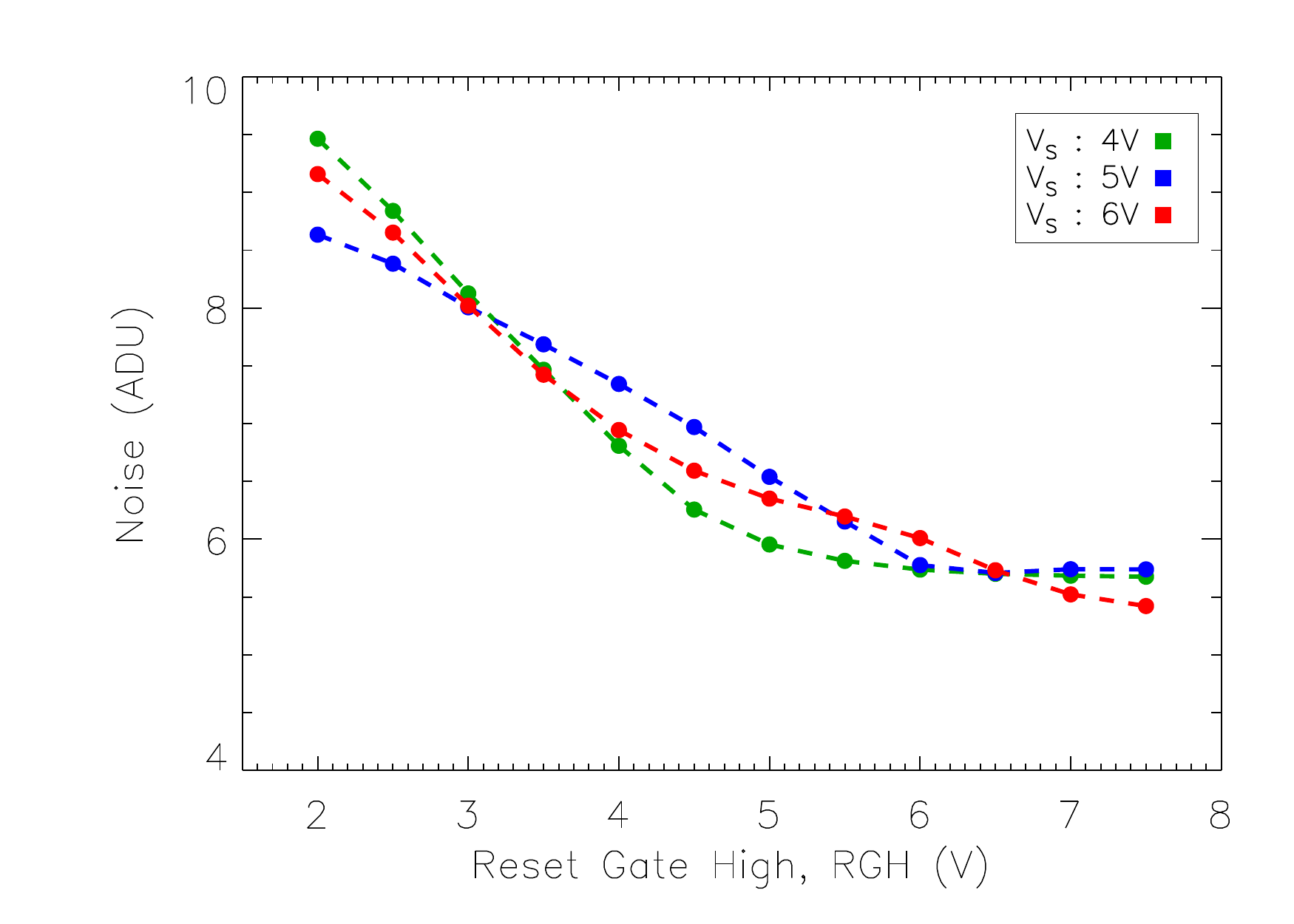}
    \caption{}
    \end{subfigure}\\
    \caption{(a) The p-MOSFET drain current as a function gate voltage, OG and RG Low bias (RGL) for three different source voltages: 4V (top), 5V (middle) and 6V (bottom). The gray dashed lines are for the 0 $\mu$A drain current level in the transistor. We optimize these parameters to find the SiSeRO operating region. (b) Optimization of the reset High bias (RGH), showing the noise with RGH for three different source voltages: 4V (green), 5V (blue) and 6V (red). See text for details.}
    \label{biasing}
\end{figure}

Characterization of the transistor was done for three different source voltages ($V_{\mathrm{S}}$): 4, 5, 6 volts, while the  source-drain ($V_{\mathrm{DS}}$) voltages were kept fixed at 5V. For each source bias, two different values of OG and RG Low, RGL (0 and 2V) were used. For each bias setting, we obtain the drain current ($I_{\mathrm{drain}}$) from the analog output of the I2V amplifier as seen from an oscilloscope (see Fig.~\ref{expt_amplifier}a), and known values of the $V_{\mathrm{S}}$, $V_{\mathrm{D}}$ and $V_{\mathrm{subtraction}}$ voltages and the bias resistors (see the schematic). Gate voltage ($V_{\mathrm{G}}$) was varied from +4V to -6V in each case. Figure \ref{biasing}a shows the drain current as a function of gate voltage for different biasing conditions. 
The top, middle and bottom panels correspond to $V_{\mathrm{S}}$ values of 4, 5, 6V respectively, while the OG and RGL potentials are shown in different colours, e.g., red (OG:2V, RGL:2V), blue (OG:2V, RGL:0V), and green (OG:0V, RGL:2V). In a p-MOSFET, we expect the drain current to start flowing when the gate bias is negative and above a certain threshold in order to form the inversion channel. However, we see that even at $V_{\mathrm{G}}$ $>$0V, there is a certain amount of drain current for some values of OG and RGL, particularly when they are low, which implies that there might be a parasitic path between the source and the drain, either along the output gate or the reset gate. Slightly positive values for OG and RGL are required to minimize the parasitic current as shown in Fig.~\ref{biasing}. We obtain the required bias for RGL and OG from the I-V characteristics of the transistor such that at $V_{\mathrm{G}}$ $>$0V, the drain current is close to zero, as shown by the gray dashed lines in Fig.~\ref{biasing}a. In Table \ref{tab1}, we give the optimum bias conditions for OG and RGL at each source voltage. Values of the serial clock potentials ($S_1$ High, $S_2$ High, $S_3$ High) depend on the bias of OG such that $S_{\mathrm{1H}}$ and $S_{\mathrm{2H}}$ and $S_{\mathrm{3H}}$ $>$ OG.    
\begin{table}[ht]
\caption{Summary of the biasing conditions of SiSeROs} 
\label{tab1}
\begin{center}       
\begin{tabular}{|l|l|l|l|} 
\hline
\rule[-1ex]{0pt}{3.5ex}  Bias potentials & $V_{\mathrm{S}}$=4V, $V_{\mathrm{D}}$=-2V & $V_{\mathrm{S}}$=5V, $V_{\mathrm{D}}$=-1V & $V_{\mathrm{S}}$=6V, $V_{\mathrm{D}}$=0V  \\
\hline\hline
\rule[-1ex]{0pt}{3.5ex}  Output Gate (OG) &+0.5V &+1V &+1.8V    \\
\hline
\rule[-1ex]{0pt}{3.5ex}  Serial clock (SL, SH) &-1, +2V &-1, +2.5V &-1, +3.5V    \\
\hline
\rule[-1ex]{0pt}{3.5ex}  Reset (RGL, RGH) &+0.5, +6.5V &+1, +6.5V &+2, +7.5V    \\
\hline
\rule[-1ex]{0pt}{3.5ex}  Reset Drain (RD) &+14V&+14V & +14V   \\
\hline
\end{tabular}
\end{center}
\end{table}

RGH (reset High) and RD should be positive with respect to the channel potential such that there is no back injection of charge to the internal channel. While RD is kept at high DC potential, with RD $>$ (RGH + channel potential), to determine the required RGH bias, we reset the internal channel soon after the charge transfer ($S_3$ is low) in the presence of X-ray photons on the detector. The CDS filtering is done between the regions before $S_3$ is low and after the second reset. At an optimum RGH bias, we do not expect to see any X-ray photons in the detector images, unless there is back injection of charge to the internal channel. To quantify this, we measure the noise in the imaging region by varying the RGH potential such that at an optimum bias setting for RGH, the noise is minimum because of minimum or no back injection of charge. This is shown Fig.~\ref{biasing}b, where we plot the estimated noise in ADU as a function of RGH for different $V_{\mathrm{S}}$ values. 
RD was kept at +14V, which is the maximum DC bias available from the Archon low voltage modules. The test was done for each source voltage while OG and RGL were kept at their respective optimum potentials. The bias condition for RGH for optimum performance of the SiSeROs is given in Table \ref{tab1}.  

\subsection{Noise and Spectral Performance}

Read noise of the system is estimated from an overclocked region of 50 $\times$ 512 pixels, by calculating the standard deviation in charge distribution of the pixels. For the centered trough SiSeRO, we estimate the read noise to be around 15 $e^{-}_{\mathrm{RMS}}$ for the optimum values of the bias parameters. In the case of the off-centered trough, the read noise is, in general, relatively higher. The best read noise that we achieved in this case is around 17 $e^{-}_{\mathrm{RMS}}$. The noise contribution from the PCB side readout electronics to these measurements is estimated to be less than 2 $e^{-}_{\mathrm{RMS}}$.  The white noise contribution of the SiSeRO is also estimated at around 2 $e^{-}_{\mathrm{RMS}}$, implying that the SiSeROs under test exhibit excess noise, possibly because the MOSFETs are surface channel and therefore are subjected to excess 1/f noise.    

\begin{figure}
    \centering
  \begin{subfigure}{.47\textwidth}
    \includegraphics[width=1\linewidth]{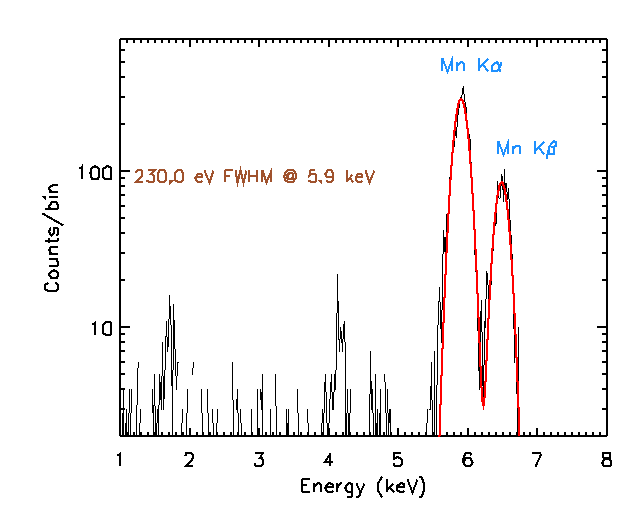}
    \caption{}
    \end{subfigure}
  \begin{subfigure}{.47\textwidth}
    \includegraphics[width=1\linewidth]{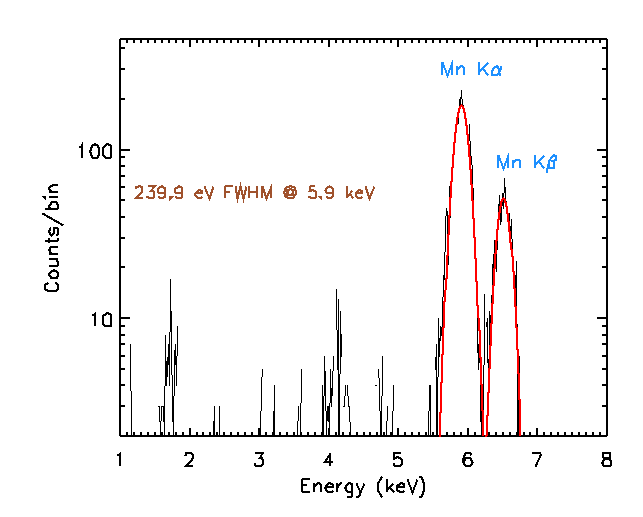}
    \caption{}
    \end{subfigure}\\
    \caption{Grade 0 spectra showing the Mn K$_{\alpha}$ (5.9 keV) and K$_{\beta}$ (6.4 keV) lines from a $^{55}$Fe radioactive source for (a) a centered trough, and (b) an off-centered trough SiSeRO at 500 kHz read out speed. The source, drain and gate voltages were kept at +6V, 0V and -10 V, respectively. (see text for further detail). The 5.9 keV and 6.4 keV lines are fitted with two Gaussian functions to estimate the gain and Full Width at Half Maximum (FWHM). The 5.9 keV FWHM values are estimated to be around 230 eV and 240 eV for the centered and off-centered SiSeRO, respectively.}
    \label{spectrum}
\end{figure}

We used an $^{55}\mathrm{Fe}$ radioisotope to evaluate the spectral performance of the detectors with 5.9 keV (Mn K$_\alpha$) and 6.4 keV (Mn K$_\beta$) X-ray photons. The X-ray images are generated after applying bias correction (subtraction of the overscan regions at 0 ms integration) and dark frame correction (subtraction of dark frames corresponding to the same time integration used for X-ray images) to the raw images. An event selection logic, based on user-defined primary and secondary pixel charge thresholds, is implemented on the X-ray images to generate graded spectra. Here we used a primary threshold of 7 times the read noise and a secondary threshold of 2.6 times the read noise.
Figures \ref{spectrum}a and  \ref{spectrum}b show examples of spectra for the centered and off-centered SiSeRO troughs, respectively. 
The spectra shown here were generated with the single pixel (grade 0) events with optimum biasing conditions of the source, drain and gate. The Mn K$_{\alpha}$ (5.9 keV) and K$_{\beta}$ (6.4 keV) lines are fitted with two Gaussian functions (shown in red solid lines) to estimate the gain and FWHM (Full Width at half Maximum). The FWHM is estimated to be around 230 eV at 5.9 keV for the centered trough SiSeRO. Because of relatively higher noise in the case of the off-centered trough, we find the FWHM to be slightly higher, namely around 240 eV. We also see an escape peak for silicon at 4.2 keV from the 5.9 keV photons. From the fitted Gaussian centroid for the 5.9 keV line and the gain of the readout module, we estimate the conversion gain of the devices to be around 700 pA per electron. 

\begin{figure}
    \centering
    \includegraphics[width=0.65\linewidth]{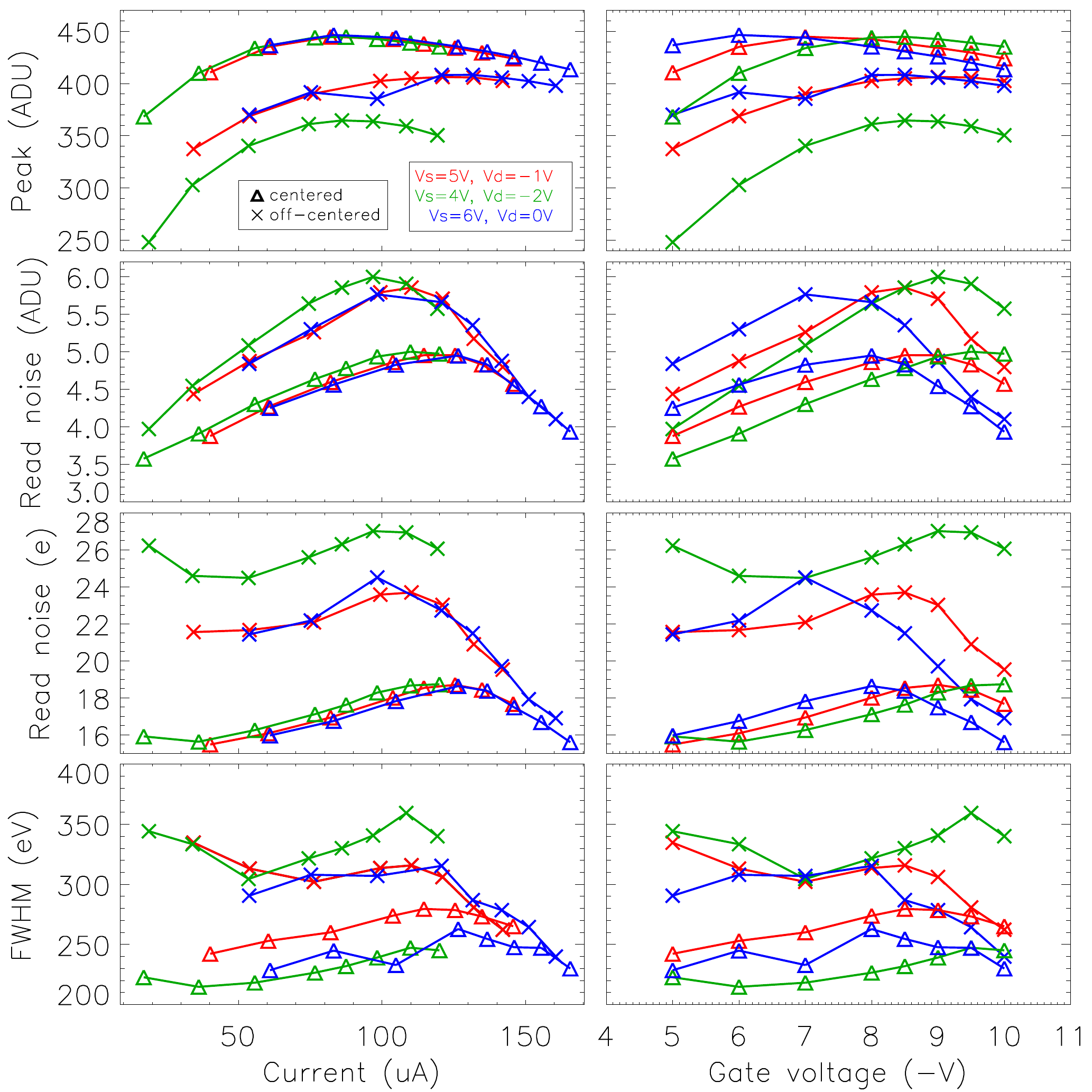}
    \caption{Spectroscopic and noise performance of the SiSeROs at various biasing conditions. Peak centroid of the 5.9 keV lines in ADU (top), read noise in ADU (second row), read noise in electron RMS (third row), and 5.9 keV line FWHM (bottom) are plotted against the p-MOSFET drain current (left column) and the corresponding gate voltages (right column). In the plots, the triangles and the crosses stand for the centered and off-centered trough, respectively. Different colors stand for various biasing conditions (refer to table \ref{tab1}).}
    \label{sisero_perf}
\end{figure}

In Fig.~\ref{sisero_perf} we summarize the spectral and noise performance of the SiSeROs for various source ($V_{\mathrm{S}}$), drain ($V_{\mathrm{D}}$) and gate ($V_{\mathrm{G}}$) voltages.    
We plot the fitted Gaussian centroid for the 5.9 keV lines in the top panel, read noise in ADU and electrons in the middle two panels, and 5.9 keV line FWHM in the bottom panel; as functions of drain current and the corresponding gate voltages in the left and right column, respectively.
The triangles and crosses stand for the centered and off-centered troughs, respectively. 
Three different source and drain voltages ($V_{\mathrm{S}}$=4V and $V_{\mathrm{D}}$=-2V, $V_{\mathrm{S}}$=5V and $V_{\mathrm{D}}$=-1V, $V_{\mathrm{S}}$=6V and $V_{\mathrm{D}}$=0V) are shown in green, red, and blue, respectively. 
We note a few interesting observations from Fig.~\ref{sisero_perf}:
\begin{itemize}
    \item Noise, gain (proportional to the centroid), and FWHM can be modeled solely with drain current, irrespective of the source and drain voltages.
    For both the variants, the gain and the output current noise initially increase with drain current, which is expected. However, when the drain current is high, the gain levels off and declines slightly, while the current noise falls strongly. The input referred read noise in electrons therefore first worsens with increased current, but then quickly improves for larger bias currents. 
    As the measured X-ray FWHM is primarily contributed by the total noise, it also follows the same trend.
    We are currently investigating this noise behavior. Potential issues could be the fact that these devices are surface channel MOSFETs and that they exhibit some parasitic leakage current around the transistor structure.
    \item The centered trough SiSeRO variant exhibits higher gain and better noise performance compared to the off-centered trough variant. In our readout, the off-centered trough is located closer to the drain, and therefore the current gain from the charge in the internal gate is reduced. For $V_{\mathrm{S}}$=4V, the gain and noise are worse for the off-centered variant, but improve with more positive source potential. This is likely due to the fact that the actual physical trough location below the MOSFET is dependent on the surrounding potentials and, for higher positive source potentials ($V_{\mathrm{S}}>$4V), the location of the most positive potential pocket moves towards the source and therefore improves the gain. Device simulations of the SiSeROs will give better understanding of the observed results.  
\end{itemize}
 
\section{Summary and Future Plans} \label{sec_future}
SiSeROs are a variant of CCDs in which signal charge can be stored within the device and moved around by changing the surface potentials. At the same time, SiSeRO devices can also serve as an output stage, enabling a broad range of innovative image sensors to be designed, including many small-area arrays, or even single-cell CCDs, where every pixel contains an individual output stage.  
In this article, we discussed the working principle of the amplifier and the concept of the drain current readout module developed at Stanford. 
We reported the first results from a prototype SiSeRO, and demonstrated that the concept is sound and can be manufactured in the single poly CCD process at MIT Lincoln Laboratory. For the optimum biasing conditions, the ENC of the prototype is around 15 electron RMS, and the FWHM at 5.9 keV is approximately 240 eV. 

In future work, we plan to mature the SiSeRO technology through a series of laboratory experiments and simulations, exploring modifications to improve the overall noise and speed performance of the detectors, e.g.,
\begin{enumerate} 
    \item The prototype SiSeRO device we produced and tested is a surface channel transistor. In this case we expect the Si/SiO$_2$ interface states to capture and release mobile carriers, causing excess noise. Therefore, we plan to test a buried transistor channel SiSeRO where the inversion layer forms some distance below the Si/SiO2 interface.  
    \item We will combine prototype measurements with 3D device simulations to improve our device modeling for a deeper understanding and improved designs. 
    \item It might be possible to utilize Repetitive Non-destructive Readout (RNDR) for SiSeROs by moving the charge packet multiple times between the internal channel and output gate. An advantage of RNDR or repetitive readout of the same charge packet is that a non-white noise (specifically 1/f or pink noise) can be significantly attenuated, resulting in extremely low noise. This technique has been successfully demonstrated for DEPFET devices with sub-electron read noise yield \cite{wolfel06}. 
    \item Parallel to our SiSeRO research and development, we are developing an application specific integrated circuit (ASIC)\cite{herrmann20_mcrc}. This will enable device readout with fewer parasitics, resulting in higher performance and offering a high channel count in a small footprint, which will be particularly useful for large, high speed X-ray imagers. The application specific design will bring a significant reduction in power consumption, while offering custom diagnostic and calibration features designed to help tune the sensors to deliver optimal performance. An 8-channel ASIC prototype designed in a 350 nm process node is currently in fabrication.   
\end{enumerate}

There is a substantial potential to utilize SiSeROs in the development of fast, low noise spectroscopic imagers for next generation X-ray astronomy missions, and a broad range of other scientific applications.


\subsection* {Acknowledgments}
This work has been supported by NASA grants APRA 80NSSC19K0499 ``Development of Integrated Readout Electronics for Next Generation X-ray CCDs", and SAT 80NSSC20K0401 ``Toward Fast, Low-Noise, Radiation-Tolerant
X-ray Imaging Arrays for Lynx: Raising Technology Readiness Further". 




\begin{thebibliography}{10}

\bibitem{Tanaka94}
Y.~{Tanaka}, H.~{Inoue}, and S.~S. {Holt}, ``{The X-Ray Astronomy Satellite
  ASCA},'' {\em Publications of the Astronomical Society of Japan} {\bf 46},
  L37--L41  (1994).

\bibitem{Lesser15_ccd}
M.~Lesser, ``A summary of charge-coupled devices for astronomy,'' {\em
  Publications of the Astronomical Society of the Pacific} {\bf 127}(957), 1097
   (2015).

\bibitem{gruner02_ccd}
S.~M. Gruner, M.~W. Tate, and E.~F. Eikenberry, ``Charge-coupled device area
  x-ray detectors,'' {\em Review of Scientific Instruments} {\bf 73}(8),
  2815--2842  (2002).

\bibitem{bautz18}
M.~Bautz, R.~Foster, B.~LaMarr, {\em et~al.}, ``{Toward fast low-noise
  low-power digital CCDs for Lynx and other high-energy astrophysics
  missions},'' in {\em Space Telescopes and Instrumentation 2018: Ultraviolet
  to Gamma Ray},  J.-W.~A. den Herder, S.~Nikzad, and K.~Nakazawa, Eds.,  {\bf
  10699}, 238 -- 248, International Society for Optics and Photonics, SPIE
  (2018).

\bibitem{bautz19}
M.~W. {Bautz}, B.~E. {Burke}, M.~{Cooper}, {\em et~al.}, ``{Toward fast,
  low-noise charge-coupled devices for Lynx},'' {\em Journal of Astronomical
  Telescopes, Instruments, and Systems} {\bf 5}, 021015  (2019).

\bibitem{bautz20}
M.~Bautz, B.~Burke, M.~Cooper, {\em et~al.}, ``{Progress toward fast,
  low-noise, low-power CCDs for Lynx and other high-energy astrophysics
  missions},'' in {\em Space Telescopes and Instrumentation 2020: Ultraviolet
  to Gamma Ray},  J.-W.~A. den Herder, S.~Nikzad, and K.~Nakazawa, Eds.,  {\bf
  11444}, 1318 -- 1323, International Society for Optics and Photonics, SPIE
  (2020).

\bibitem{matsunaga91}
Y.~Matsunaga, H.~Yamashita, and S.~Ohsawa, ``A highly sensitive on-chip charge
  detector for ccd area image sensor,'' {\em IEEE Journal of Solid-State
  Circuits} {\bf 26}(4), 652--656  (1991).

\bibitem{kemmer87_depfet}
J.~Kemmer and G.~Lutz, ``New detector concepts,'' {\em Nuclear Instruments and
  Methods in Physics Research Section A: Accelerators, Spectrometers, Detectors
  and Associated Equipment} {\bf 253}(3), 365--377  (1987).

\bibitem{strueder00_depfet_imager}
L.~Strueder, N.~Meidinger, E.~Pfeffermann, {\em et~al.}, ``{Fully depleted
  backside-illuminated spectroscopic active pixel sensors from the infrared to
  x rays (1 eV to 25 keV)},'' in {\em X-Ray Optics, Instruments, and Missions
  III},  J.~E. Truemper and B.~Aschenbach, Eds.,  {\bf 4012}, 200 -- 217,
  International Society for Optics and Photonics, SPIE  (2000).

\bibitem{treberspurg20_wfi}
W.~{Treberspurg}, R.~{Andritschke}, A.~{Behrens}, {\em et~al.},
  ``{Characterization of a 256 {\texttimes} 256 pixel DEPFET detector for the
  WFI of Athena},'' {\em Nuclear Instruments and Methods in Physics Research A}
  {\bf 958}, 162555  (2020).

\bibitem{Chattopadhyay20_spie}
T.~Chattopadhyay, S.~Herrmann, S.~Allen, {\em et~al.}, ``{Tiny-box: a tool for
  the versatile development and characterization of low noise fast x-ray
  imaging detectors},'' in {\em X-Ray, Optical, and Infrared Detectors for
  Astronomy IX},  A.~D. Holland and J.~Beletic, Eds.,  {\bf 11454}, 368 -- 385,
  International Society for Optics and Photonics, SPIE  (2020).

\bibitem{archon14}
G.~Bredthauer, ``{Archon: A modern controller for high performance astronomical
  CCDs},'' in {\em Ground-based and Airborne Instrumentation for Astronomy V},
  S.~K. Ramsay, I.~S. McLean, and H.~Takami, Eds.,  {\bf 9147}, 1730 -- 1740,
  International Society for Optics and Photonics, SPIE  (2014).

\bibitem{wolfel06}
S.~Wölfel, S.~Herrmann, P.~Lechner, {\em et~al.}, ``Sub-electron noise
  measurements on repetitive non-destructive readout devices,'' {\em Nuclear
  Instruments and Methods in Physics Research Section A: Accelerators,
  Spectrometers, Detectors and Associated Equipment} {\bf 566}(2), 536--539
  (2006).

\bibitem{herrmann20_mcrc}
S.~Herrmann, J.~Wong, T.~Chattopadhyay, {\em et~al.}, ``{MCRC V1: development
  of integrated readout electronics for next generation x-ray CCD detectors for
  future satellite observatories},'' in {\em X-Ray, Optical, and Infrared
  Detectors for Astronomy IX},  A.~D. Holland and J.~Beletic, Eds.,  {\bf
  11454}, 412 -- 418, International Society for Optics and Photonics, SPIE
  (2020).

\end{thebibliography}




\end{spacing}
\end{document}